\documentclass[floatfix,lengthcheck,showpacs,amssymb,amsmath,amsfonts,twocolumn]{aastex63}

\usepackage{lineno}
\usepackage[utf8]{inputenc}
\usepackage{color}
\usepackage{graphicx}
\usepackage{float}
\usepackage{subfigure}
\usepackage{amsmath}
\usepackage{afterpage}
\usepackage{acronym}
\usepackage{multirow}
\usepackage{tikz}
\usetikzlibrary{arrows,shapes,trees,decorations.pathreplacing}
\usepackage{import}
\usepackage{svn-multi}
\usepackage{lineno}
\usepackage{hyperref}
\usepackage{gensymb}
\hypersetup{
    colorlinks=true,
    linkcolor=blue,
    filecolor=magenta,      
    urlcolor=cyan,
}

\newcommand{\msun}{\ensuremath{\mathrm{M}_{\odot}}}

\begin{document}
\title[]{GW190521 may be an intermediate mass ratio inspiral}

\correspondingauthor{Alexander H. Nitz}
\email{alex.nitz@aei.mpg.de}

\author[0000-0002-1850-4587]{Alexander H. Nitz}
\author[0000-0002-0355-5998]{Collin D. Capano}
\affil{Max-Planck-Institut f{\"u}r Gravitationsphysik (Albert-Einstein-Institut), D-30167 Hannover, Germany}
\affil{Leibniz Universit{\"a}t Hannover, D-30167 Hannover, Germany}


\begin{abstract}
GW190521 is the first confident observation of a binary black hole merger with total mass $M > 100\,\mathrm{M}_{\odot}$. Given the lack of observational constraints at these masses, we analyze GW190521 considering two different priors for the binary's masses: uniform in mass ratio and source-frame total mass, and uniform in source-frame component masses. For the uniform in mass-ratio prior, we find that the component masses are $m_1^{\mathrm{src}} = 168_{-61}^{+15}\,\mathrm{M}_{\odot}$ and $m_2^{\mathrm{src}} = 16_{-3}^{+33}\,\mathrm{M}_{\odot}$. The uniform in component-mass prior yields a bimodal posterior distribution. There is a low-mass-ratio mode ($q<4$) with $m_1^{\mathrm{src}} = 100_{-18}^{+17}\,\mathrm{M}_{\odot}$ and $m_2^{\mathrm{src}} = 57_{-16}^{+17}\,\mathrm{M}_{\odot}$ and a high-mass-ratio mode ($q\geq4$) with $m_1^{\mathrm{src}} = 166_{-35}^{+16}\,\mathrm{M}_{\odot}$ and $m_2^{\mathrm{src}} = 16_{-3}^{+14}\,\mathrm{M}_{\odot}$. Although the two modes have nearly equal posterior probability, the maximum-likelihood parameters are in the high-mass ratio mode, with $m_1^{\rm src} = 171\,M_{\odot}$ and $m_2^{\rm src} = 16\,M_{\odot}$, and a signal-to-noise ratio of $16$. These results are consistent with the proposed ``mass gap'' produced by pair-instability in supernova. Our results differ from those published in~\cite{Abbott:2020tfl}. We find that a combination of the prior used and the constraints applied may have prevented that analysis from sampling the high-mass-ratio mode. An accretion flare in AGN J124942.3+344929 was observed in possible coincidence with GW190521 by the Zwicky Transient Facility (ZTF). We report parameters assuming a common origin; however, the spatial agreement of GW190521 and the EM flare alone does not provide convincing evidence for the association ($\ln\mathcal{B} \gtrsim -4$).

\end{abstract}

\section{Introduction}
Gravitational-wave astronomy began with the observation of GW150914~\citep{Abbott:2016blz} by the twin LIGO-Hanford and Livingston observatories~\citep{TheLIGOScientific:2014jea} with the merger of two $\sim 30\msun$ black holes, significantly heavier than previously known black holes in X-ray binaries~\citep{Corral-Santana:2015fud}. These heavy binary black holes (BBHs) opened a new window into stellar evolution~\citep{PhysRevD.98.083017,Dvorkin:2017kfg,Piran:2018bbt} and even sparked renewed
interest in primordial black holes as a component of dark matter~\citep{Green:2020jor,Nitz:2020bdb,Authors:2019qbw}. Since then, the Virgo observatory~\citep{TheVirgo:2014hva} has joined the growing worldwide observatory network and
over a dozen binary black hole mergers have been observed~\citep{Nitz:2018imz,Nitz:2019hdf,Nitz:2020naa,Venumadhav:2019tad,Venumadhav:2019lyq,Zackay:2019btq,LIGOScientific:2018mvr}, with many additional candidates awaiting publication~\citep{GraceDBo3page}. 

With the exception of the marginal BBH candidates GW151205 and 170817+03:02:46UTC~\citep{Nitz:2019hdf,Zackay:2019btq}, all prior confident detections were consistent with sources in which both component black holes have mass less than $50\msun$~\citep{LIGOScientific:2018jsj}. This observed limit may hint at the existence of an upper mass gap~\citep{Fishbach:2017zga, Talbot:2018cva,LIGOScientific:2018jsj,Roulet:2020wyq}. Formation models which include the effects of pulsational pair instability supernovae (PPISNe) or pair-instability supernovae (PISNe) in stellar evolution preclude the direct formation of a black hole with remnant mass $\sim50$--$120\,\msun$~\citep{2016MNRAS.457..351Y,Woosley:2016hmi,Belczynski:2016jno,Marchant:2018kun,Woosley_2019,Stevenson:2019rcw,vanSon:2020zbk}.

On May 21st, 2019 at 03:02:29 UTC, GW190521 was initially reported by the PyCBC Live low-latency analysis~\citep{Nitz:2018rgo,DalCanton:2020vpm}, in addition to detection by the cWB~\citep{Klimenko:2015ypf}, gstLAL~\citep{Messick:2016aqy}, and SPIIR~\citep{chu:2020pjv} analyses, producing a 765 deg$^2$ Bayestar sky localization~\citep{BAYESTAR}. Continued monitoring of the low-latency localization region was conducted by ZTF~\citep{2019PASP..131a8002B}, which detected a flare weeks later that was consistent with AGN J124942.3+344929 (ZTF19abanrhr)~\citep{Graham:2020gwr} at $z=0.438$ from the Million Quasar Catalog~\citep{2019arXiv191205614F}. If there is a common origin for the gravitational-wave and EM flare, it would suggest that the binary merger occurred within the accretion disk of an active galactic nuclei. 

The initial analysis of GW190521 by the LIGO and Virgo Collaborations (LVC) estimated that it had a primary $m_1^{\mathrm{src}}$ and secondary mass $m_2^{\mathrm{src}}$ of $85_{-14}^{+21}\,M_{\odot}$ and $66_{-18}^{+17}\,M_{\odot}$, respectively~\citep{Abbott:2020tfl}. This would mean one or both of the black holes lie within the PISN mass gap. If true, GW190521 would either challenge existing stellar formation theory, or provide the first instance of a new class of hierarchically formed binaries. Several papers, including the original announcement, explored this and other possibilities~\citep{Abbott:2020tfl,DeLuca:2020sae,Kremer:2020wtp,Costa:2020xbc,Ziegler:2020klg,Safarzadeh:2020vbv, Belczynski:2020bca,Fragione:2020han}.

\setlength{\tabcolsep}{1pt}
\begin{table}[ht!]
  \begin{center}
    \caption{The 90\% credible marginal intervals for GW190521 using the IMRPhenomXPHM model. In order, the intervals for the gravitational-wave only prior using either a uniform in mass ratio $q$ and source-frame total mass $M^{\mathrm{src}}$ prior, uniform in source-frame component masses $m_{1,2}$ prior, and the maximum-likelihood parameters. We show the marginal intervals separately for both the low-q ($q<4$) and high-q ($q>4$) modes evident when using a uniform in $m_{1,2}^{\mathrm{src}}$ prior. On the right are the intervals when we limit to the observed location of the EM flare for the uniform in q-M prior.}
    \label{table:params}
    
\begin{tabular}{c|c|c|c|c|c}
& \multicolumn{4}{|c|}{GW-only}& GW+EM \\ \hline
Parameter & $q$-$M^{\rm src}$ prior & \multicolumn{2}{|c|}{$m_{1,2}^{\mathrm{src}}$ prior} & ML & $q$-$M^{\rm src}$ prior \\ \hline
&&low-$q$&high-$q$&&\\
$m_1^{\mathrm{src}}\, [\msun]$& $168_{-61}^{+15}$ & $100_{-18}^{+17}$ & $166_{-35}^{+16}$ & 171 &  $108_{-12}^{+33}$ \\&&&&& \\
$m_2^{\mathrm{src}}\, [\msun]$& $16_{-3}^{+33}$ & $57_{-16}^{+17}$ & $16_{-3}^{+14}$ & 16& $47_{-23}^{+22}$ \\&&&&&\\
$M^{\mathrm{src}}\, [\msun]$& $184_{-30}^{+15}$ & $156_{-15}^{+21}$ & $183_{-27}^{+15}$ & 187& $156_{-14}^{+17}$ \\&&&&&\\
$q$ & $10.7_{-8.6}^{+2.4}$ & $1.8_{-0.6}^{+1.0}$ & $10.3_{-5.7}^{+2.4}$ &10.6 & $2.2_{-0.8}^{+3.3}$\\&&&&&\\
$\chi_{\mathrm{eff}}$ & $-0.51_{-0.11}^{+0.24}$ & $-0.16_{-0.40}^{+0.42}$ & $-0.53_{-0.12}^{+0.14}$ & -0.55& $-0.45_{-0.23}^{+0.40}$\\&&&&&\\
$|\chi_1|$ & $0.85_{-0.25}^{+0.11}$ & $0.72_{-0.59}^{+0.25}$ & $0.87_{-0.16}^{+0.10}$ & 0.89 & $0.92_{-0.28}^{+0.06}$ \\&&&&&\\
$d_L\,[\mathrm{Mpc}]$ & $1060_{-280}^{1400} $ & $3130_{-1500}^{+2260}$ & $1100_{-310}^{+900}$ & 950 & -  \\

\end{tabular}
  \end{center}
\end{table} 

In this letter, we analyze GW190521 using a standard astrophysical prior that the rate is uniform in comoving volume. Current estimates for the mass ratio distribution from the gravitational-wave observation of lighter stellar-mass black hole mergers are still weakly constrained~\citep{GW190412}; given that GW190521's total mass is at the boundary of the observed binaries, it may be the first instance of a separate population. In light of this, we consider two priors for the binary's masses: (1) flat in mass ratio $q$ from 1-25 (where we define $q \geq 1$; i.e., as the ratio of the larger mass to the smaller mass) and source-frame total mass $M^{\rm src}$ from 80-300\msun, and (2) uniform in source-frame component masses $m^{\rm src}_{1,2}$. We also separately combine the distance and sky location information from the electromagnetic observation under the assumption of a common origin.

We find that GW190521 has posterior support for larger mass ratio than previously reported under both mass priors. With the uniform in mass ratio prior, $93\%$ of the posterior is at $q > 4$. The uniform in component mass prior yields a bimodal posterior in mass ratio, with $51\%$ of the posterior has $q > 4$. The component masses are consistent with neither component lying within the mass gap from pair-instability supernova. This is consistent with the interpretation reached in~\cite{Fishbach:2020qag}, and along similar reasoning as proposed in~\cite{Fishbach:2020qag}, if one considers even a moderate prior for the existence of a mass gap, then the high-mass-ratio mode is selected for at high confidence. This could make it the first intermediate-mass-ratio inspiral (IMRI) observed~\citep{Brown:2006pj}. The LIGO and Virgo Collaborations (LVC) also used a prior uniform in component mass in their analysis of GW190521, but did not find the high-mass-ratio mode in the posterior~\citep{Abbott:2020tfl}. We find that this discrepancy is likely due to a combination of constraints used in the LVC analysis that excluded the highest likelihood regions, along with a difference in distance prior, and the overall complexity of sampling the parameter space.

\begin{figure*}[ht!]
    \includegraphics[width=2.0\columnwidth]{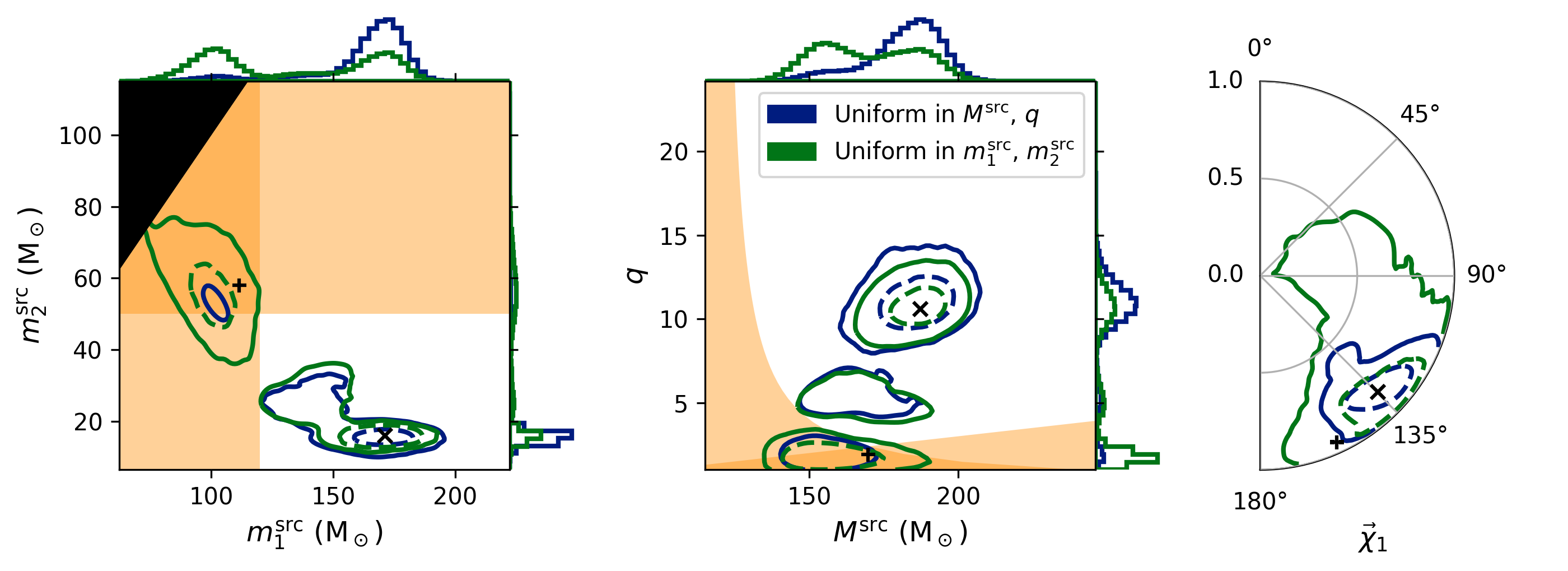}
    \caption{Posterior distribution for GW190521 using a prior uniform in mass ratio $q$ and source-frame total mass $M_{\mathrm{src}}$ (blue) and using a prior uniform in source-frame component masses $m_{1,2}^{\mathrm{src}}$ (green). In both cases a prior uniform in comoving volume and isotropic in sky location is used. Shown are the source-frame component masses $m_{1,2}^{\rm src}$ (left), mass ratio $q$ versus source-frame total mass $M^{\rm src}$ (center), and dimensionless spin of the more massive black hole $\vec{\chi}_{1}$ (right). The $\vec{\chi}_{1}$ angle is measured with respect to the orbital angular momentum at a fiducial gravitational-wave frequency of $20\,$Hz, with $0^\circ$ corresponding to aligned spin. Orange regions indicate parts of parameter space where one or both component objects are in the ``mass gap'' $[50, 120)\,\msun$. Dashed and solid contours indicate the 50\% and 90\% credible regions, respectively. Contours are drawn by applying a multimodal kernel density estimate~\citep{kombine} to the samples. The one dimensional marginal distributions are shown along the axes for the left and middle panels. The black plus and cross in the left and middle plots are the maximum-likelihood points in the regions $q \in [1,2)$ and $q \in [2, 25)$, respectively. The gravitational waveform corresponding to each of these points is shown in Fig.~\ref{fig:waveform}.}
\label{fig:mprior}
\end{figure*}

\section{Gravitational-wave Inference}

We estimate the source parameters of GW190521 using the open-source PyCBC Inference library~\citep{Biwer:2018osg,pycbc-github} along with the Dynesty nested sampler~\citep{speagle:2019} and the public LIGO and Virgo data~\citep{Vallisneri:2014vxa,Abbott:2019ebz}. A low and high frequency cutoff at 20 and 500 Hz, respectively, are used for the evaluation of the likelihood. As a necessary component to calculate the likelihood, we estimate the power spectral density of the data using a Welch method, where we do a median-mean average over 71 8-second segments, centered on the merger time.

We empirically find that the likelihood surface for GW190521 has a complex multi-modal structure. To obtain an accurate estimate of the posterior density, we found it necessary to numerically marginalize over the polarization angle $\psi$ and the coalescence phase $\phi_c$ using a $1000\times240$ grid. These parameters are strongly coupled to each other and the inclination angle $\iota$ (here defined as the angle between the orbital angular momentum and the line-of-sight at a fiducial gravitational-wave frequency of $20\,$Hz) due to the the relatively few observable cycles in GW190521. Typical gravitational-wave analyses have found self-consistent posteriors using Dynesty with 3--5000 live points~\citep{Romero-Shaw:2020owr}. However, to ensure sampling of the parameter space with separated modes that occupy vastly different volumes of the prior space, we conduct our analysis with $20$--$40\,000$ live points. This helps guard against mode ``die-off'' whereby the sampler neglects entire modes if they are not populated by live points early enough in the analysis.

\begin{figure*}[ht!]
    \includegraphics[width=2.0\columnwidth]{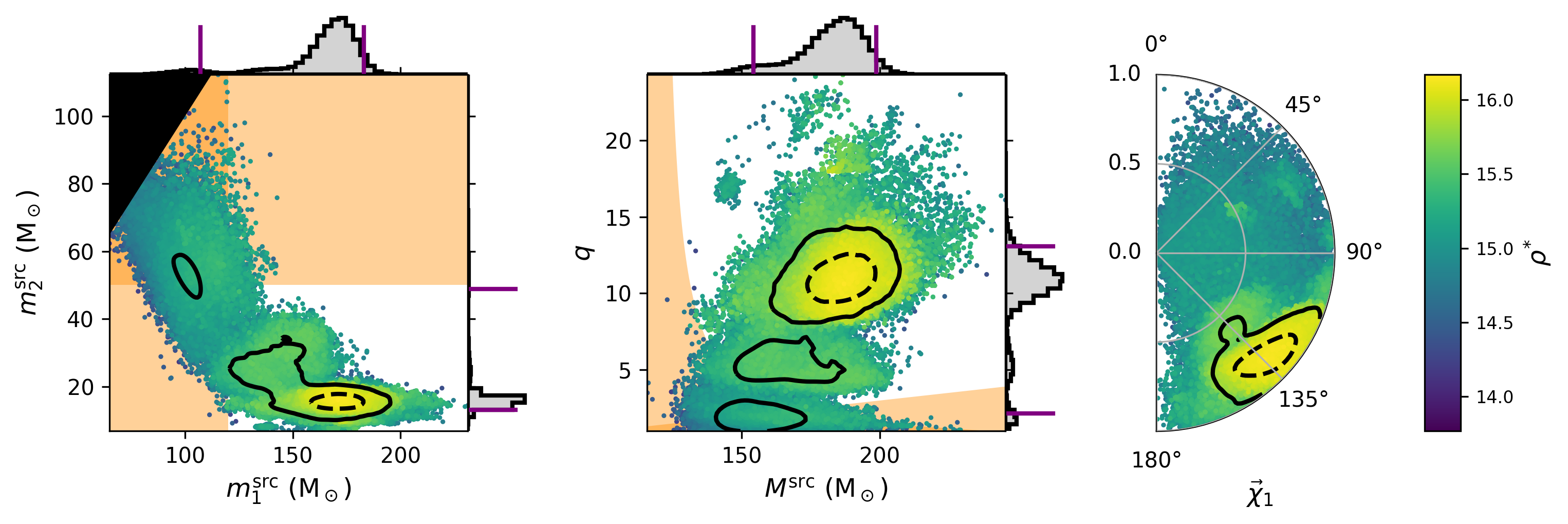} \\
    \includegraphics[width=2.0\columnwidth]{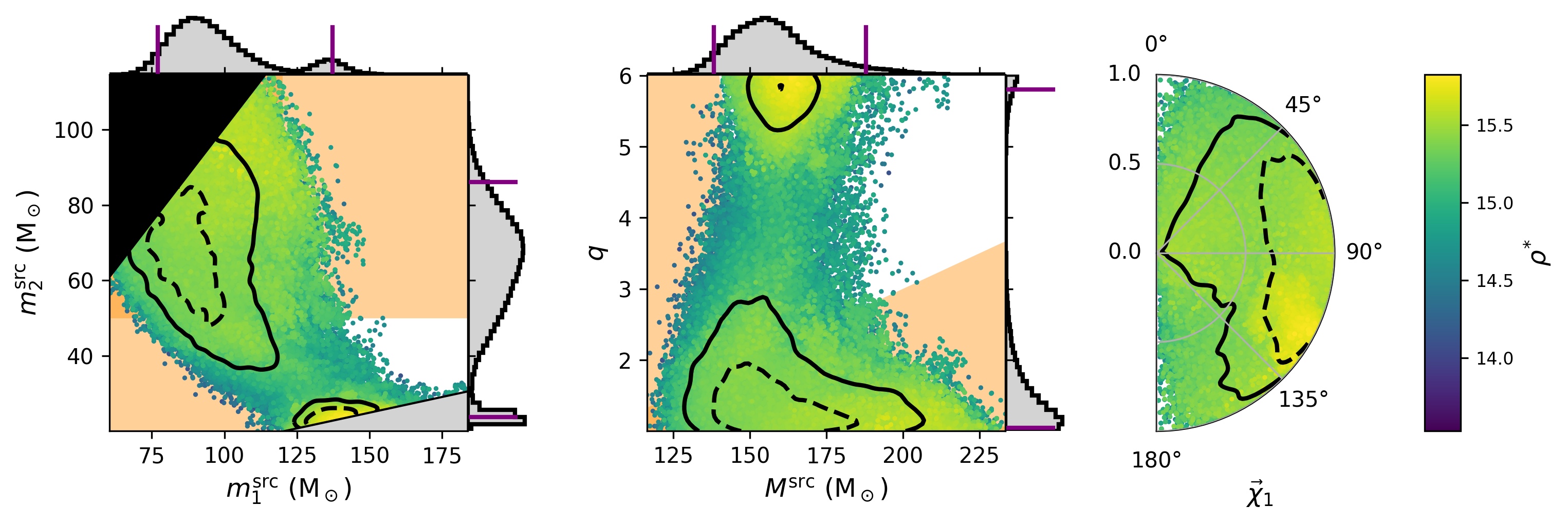}
    \caption{Posterior distribution for GW190521 with a uniform prior in mass ratio and source-frame total mass, as well as uniform in comoving volume and isotropic in sky location. Results for IMRPhenomXPHM (top) and NRSur7dq4 (bottom) are shown. Parameters shown, and the definition of contours and orange regions are the same as in Fig.~\ref{fig:mprior}. The gray region in the bottom component mass plot indicates the area of parameter space excluded by the NRSur7dq4 model. Posterior samples are colored by their signal-to-noise ratio $\rho\equiv(2\ln\mathcal{L})^{1/2}$, where $\mathcal{L}$ is the likelihood ratio at the given point. The one dimensional marginal distributions are shown along the axes for the left and middle panels along with the 90\% credible interval.}
\label{fig:ref}
\end{figure*}

We consider two priors for the black hole masses: a prior that is flat in the source-frame total mass $M^{\rm src}$ between $80-300\msun$ and is uniform in the mass ratio q from $1-25$, along with a prior which is uniform in component masses within the same bounds. We achieve the latter posterior by re-weighting our analysis which used a uniform in mass ratio prior. We also separately consider the possibility that the EM flare and GW merger share a common origin. In all cases, we use a prior isotropic in orientation and uniform in magnitude between $0$ and $0.99$ for each object's dimensionless spin $\vec{\chi}_{1,2}$. The binary orientation given by its inclination $\iota$ and polarization angle $\psi$ is chosen to be isotropic. In the case where we are agnostic to a possible common origin for the observed flare and gravitational-wave merger, we use a prior that is uniform in comoving volume and isotropic in sky location. When assuming a common origin for the GW and flare observations, we use the sky location and redshift of the flare to fix the source location, where the luminosity distance $d_L$ is determined from redshift using a standard $\Lambda$CDM cosmology\citep{Ade:2015xua}.

To model the gravitational waveform we use the recently developed IMRPhenomXPHM model~\citep{Pratten:2020ceb}. This waveform models a quasi-circular BBH merger including sub-dominant harmonics and the effects of precession. IMRPhenomXPHM includes improvements over the IMRPhenomPv3HM model~\citep{Khan:2019kot}, which is one of the models used in the original analysis of GW190521~\citep{Abbott:2020tfl}. These improvements include calibration of the sub-dominant harmonics against numerical relativity simulations and high-mass-ratio waveforms produced through perturbative expansions of general relativity. 

For comparison to IMRPhenomXPHM, we also analyze GW190521 with NRSur7dq4~\citep{Varma:2019csw}, which is a model based on the interpolation of numerical relativity simulations up to $q=4$ and includes an extrapolation up to $q=6$. Due to the limited range of the model, we cannot perform a full comparison to the results of IMRPhenomXPHM. However, we can use NRSur7dq4 to crosscheck IMRPhenomXPHM within the models' common regions of validity. We note that due to the lack of full numerical relativity simulations with near extremal spin at the highest mass ratios in our prior, the IMRPhenomXPHM model also represents a kind of extrapolation. Additional simulations in this region would help mitigate any potential systematics.

\begin{figure*}[ht!]
    \includegraphics[width=2.0\columnwidth]{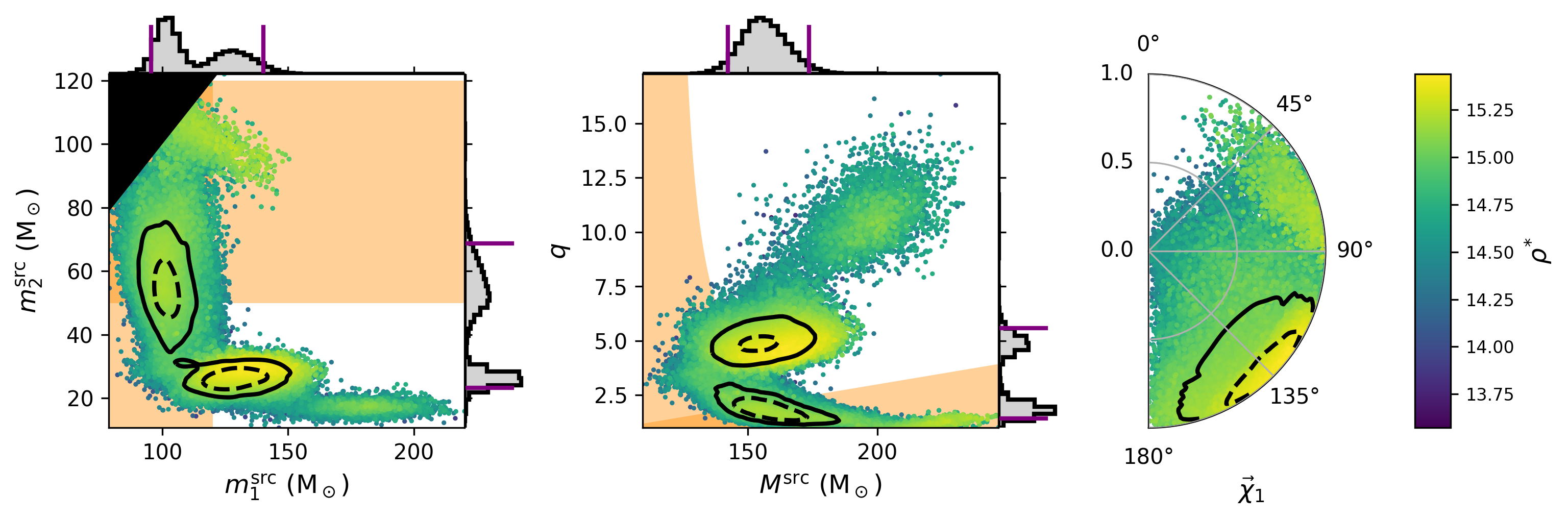} \\
    \includegraphics[width=2.0\columnwidth]{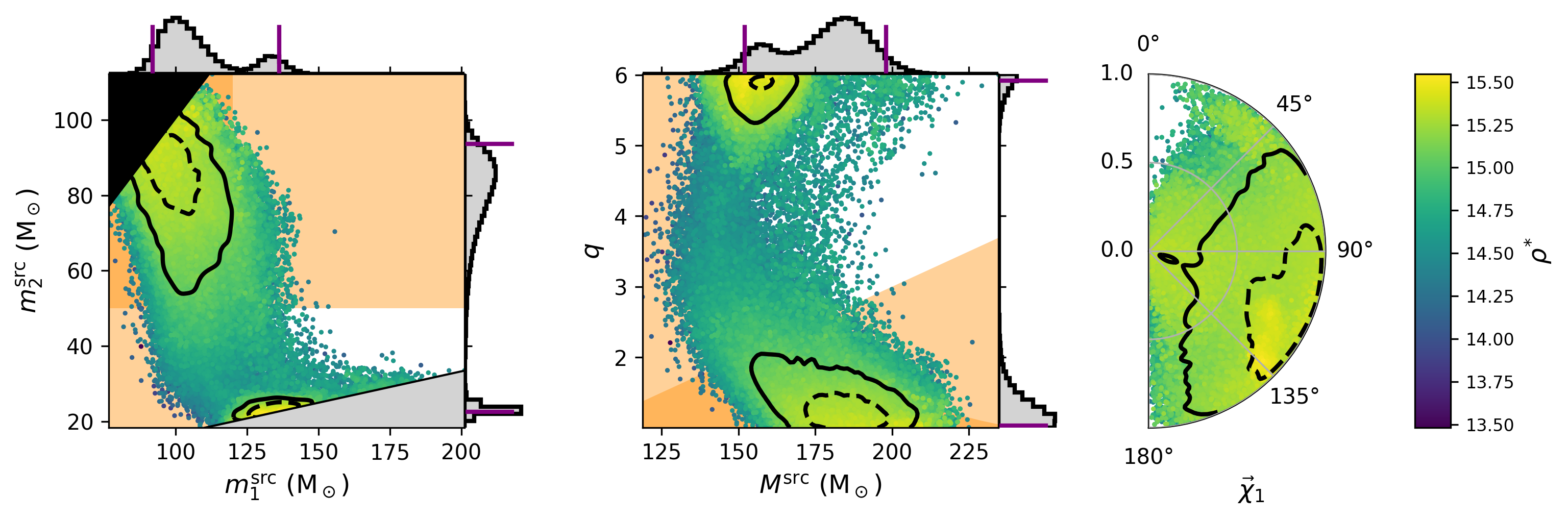}
    \caption{Posterior distribution for GW190521 using the location of the EM flare observed by ZTF to fix the sky location and redshift. Results for IMRPhenomXPHM (top) and NRSur7dq4 (bottom) are shown. See Fig.~\ref{fig:ref} for plot details.}
\label{fig:com}
\end{figure*}

\begin{figure}[ht!]
    \includegraphics[width=\columnwidth]{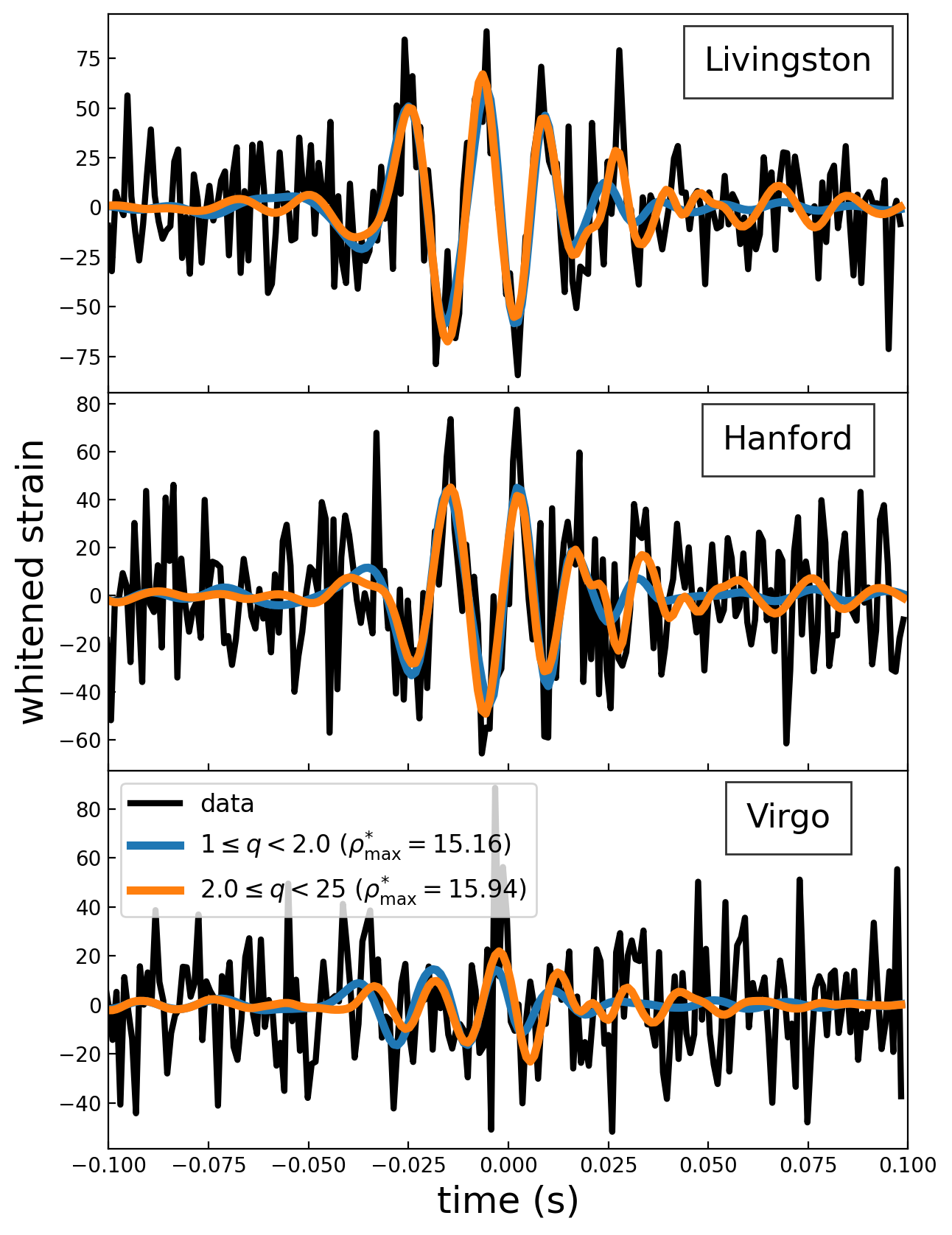}
    \caption{Whitened data and gravitational waveform model as seen in each detector. Shown are the maximum-likelihood IMRPhenomXPHM waveforms in the regions $q \in [1,2)$ and $q \in [2, 25)$ (corresponding to the plus, and cross in the top plots of Fig.~\ref{fig:mprior}, respectively). These have network signal-to-noise ratios of 15.39 and 16.17, respectively. The data and waveforms are time shifted such that $t=0$ corresponds to the the coalescence time of the maximum-likelihood waveform in each detector.}
\label{fig:waveform}
\end{figure}
\vspace{21pt}
\section{Results}

The posterior results for GW190521 using just the gravitational-wave data and both mass priors are shown in Fig.~\ref{fig:mprior}. Figure~\ref{fig:ref} shows a comparison of the IMRphenomXPHM and NRSur7dq4 models using the uniform in mass ratio prior. Figure~\ref{fig:com} shows the posterior results when assuming a common source with the observed EM flare. The credible intervals for the IMRPhenomXPHM model are summarized in Table~\ref{table:params}. Notably, all posteriors exhibit some level of multi-modality. 

For the GW-only posteriors, we find that, within their common range of mass ratio ($q < 6$), both the IMRPhenomXPHM and NRSur7dq4 models give comparable results. In both cases, the maximum-likelihood point is at mass ratios $> 5$. The NRSur7dq4 model is clearly limited by the inability to sample above $q=6$. For the IMRPhenomXPHM model (which we allow to explore up to $q=25$) we find that the highest likelihood is at $q\sim11$.

When we constrain the sky location and redshift to those of the observed flare, we find that the two modes at $q\sim 1-2$ and $q\sim5-6$ have approximately equal posterior support, but the highest mass ratio mode is nearly excluded. As there is a degeneracy between the mass ratio and distance --- higher mass ratios imply a closer source --- the substantial difference in preferred mass ratio arises from the tight distance constraint of the flare. The reality of the flare association will have a substantial impact on the understanding of GW190521's source parameters. 

\begin{figure*}[ht!]
    \includegraphics[width=\columnwidth]{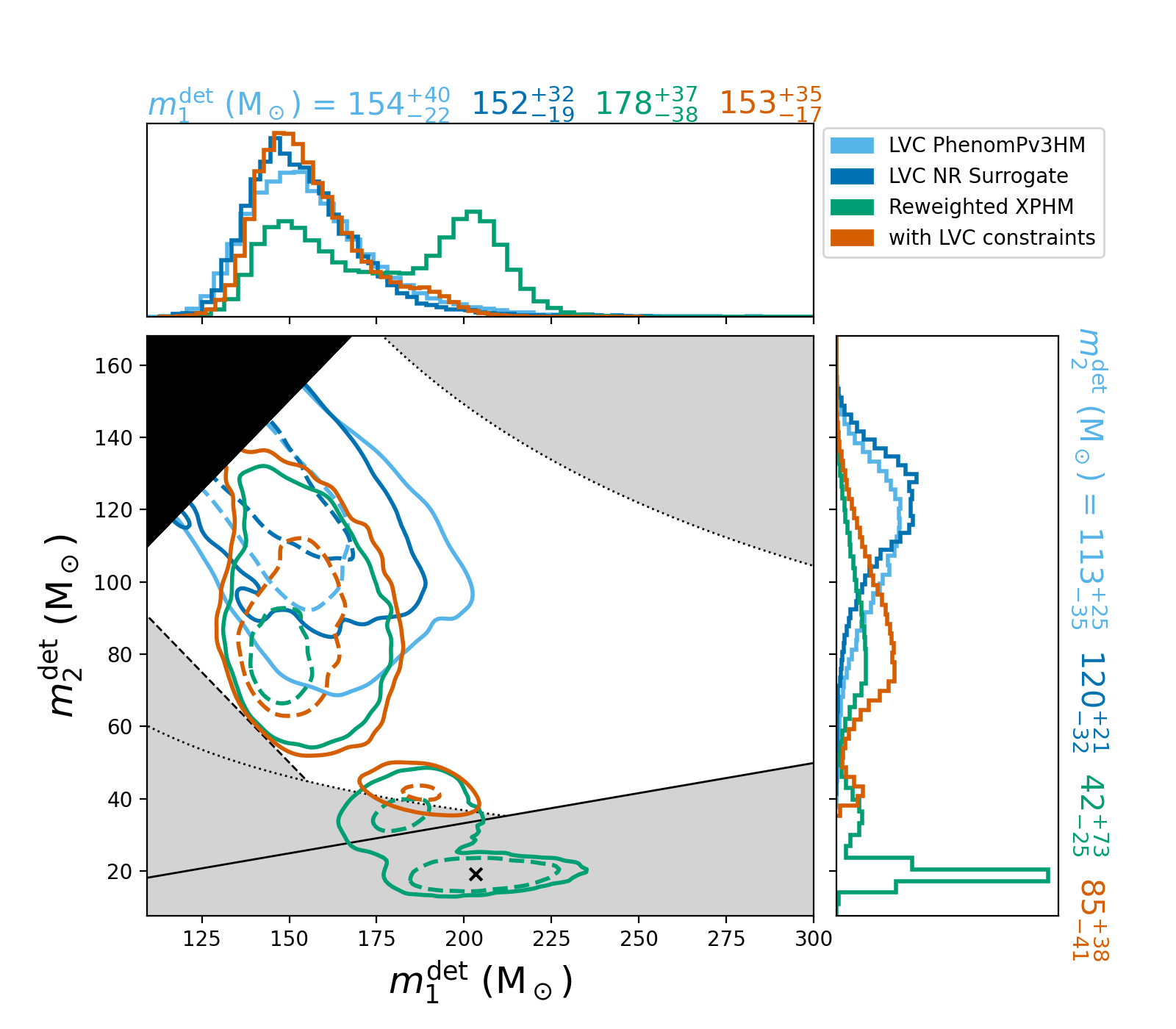}
    \includegraphics[width=\columnwidth]{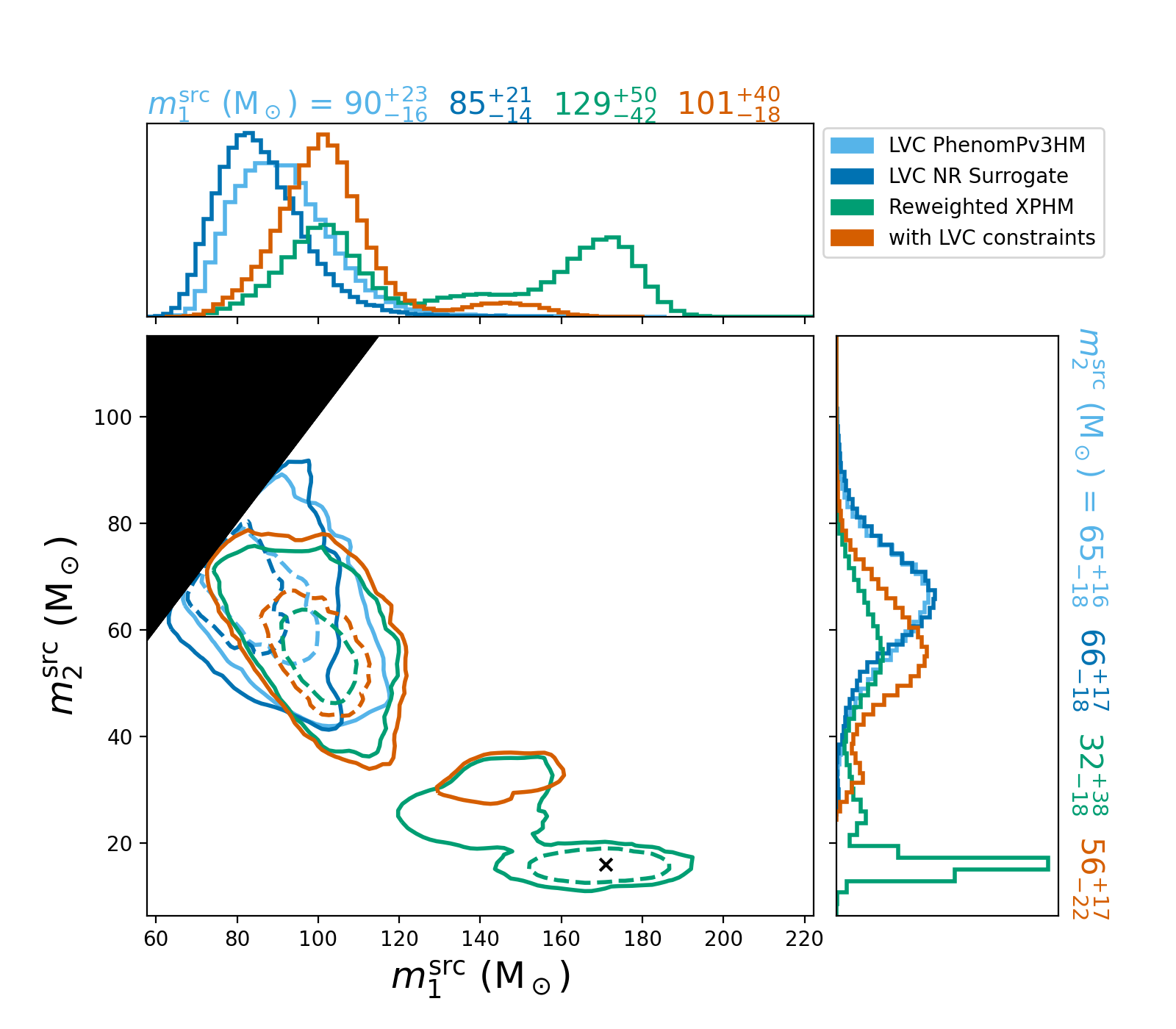}\\
    \includegraphics[width=\columnwidth]{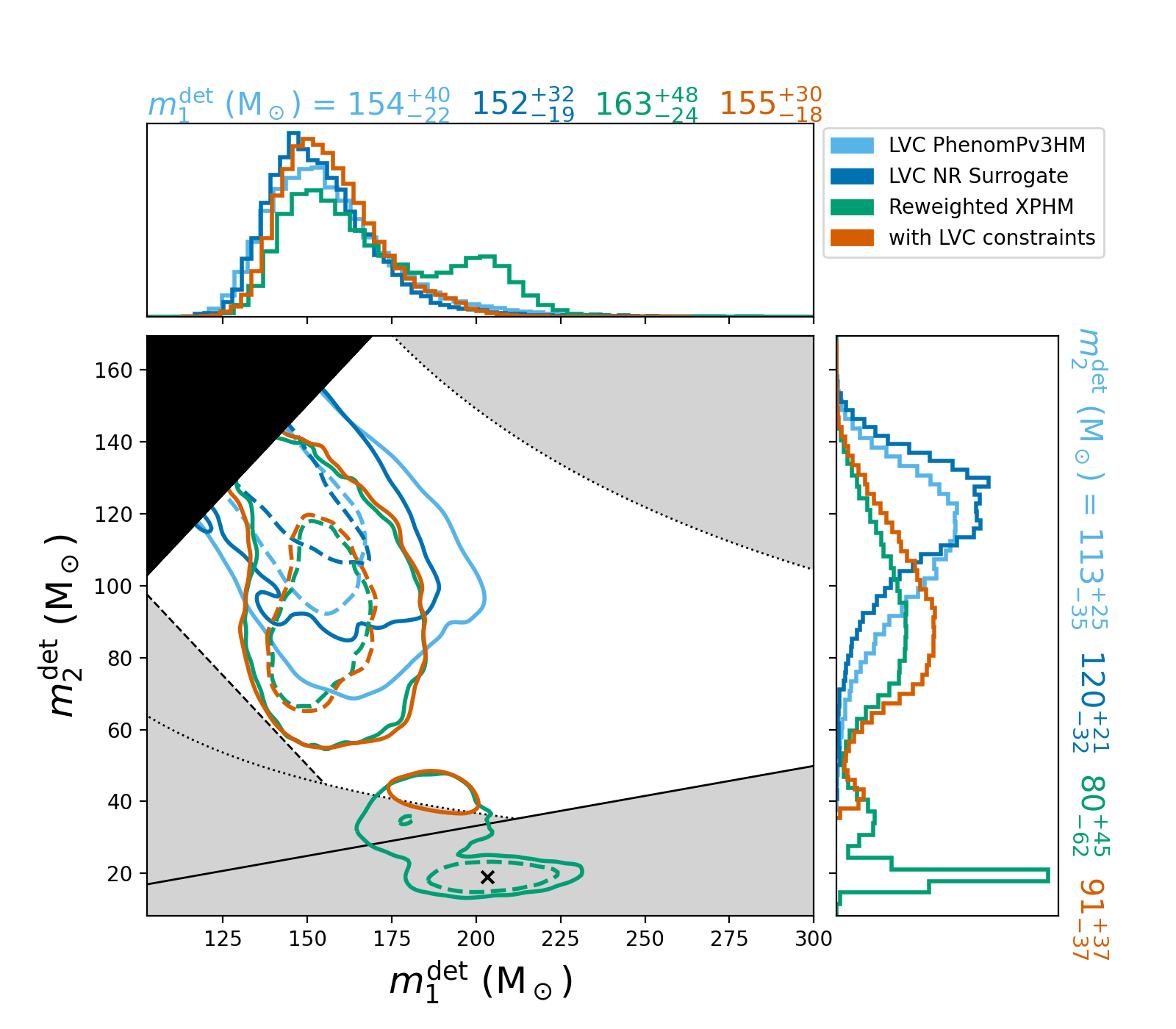}
    \includegraphics[width=\columnwidth]{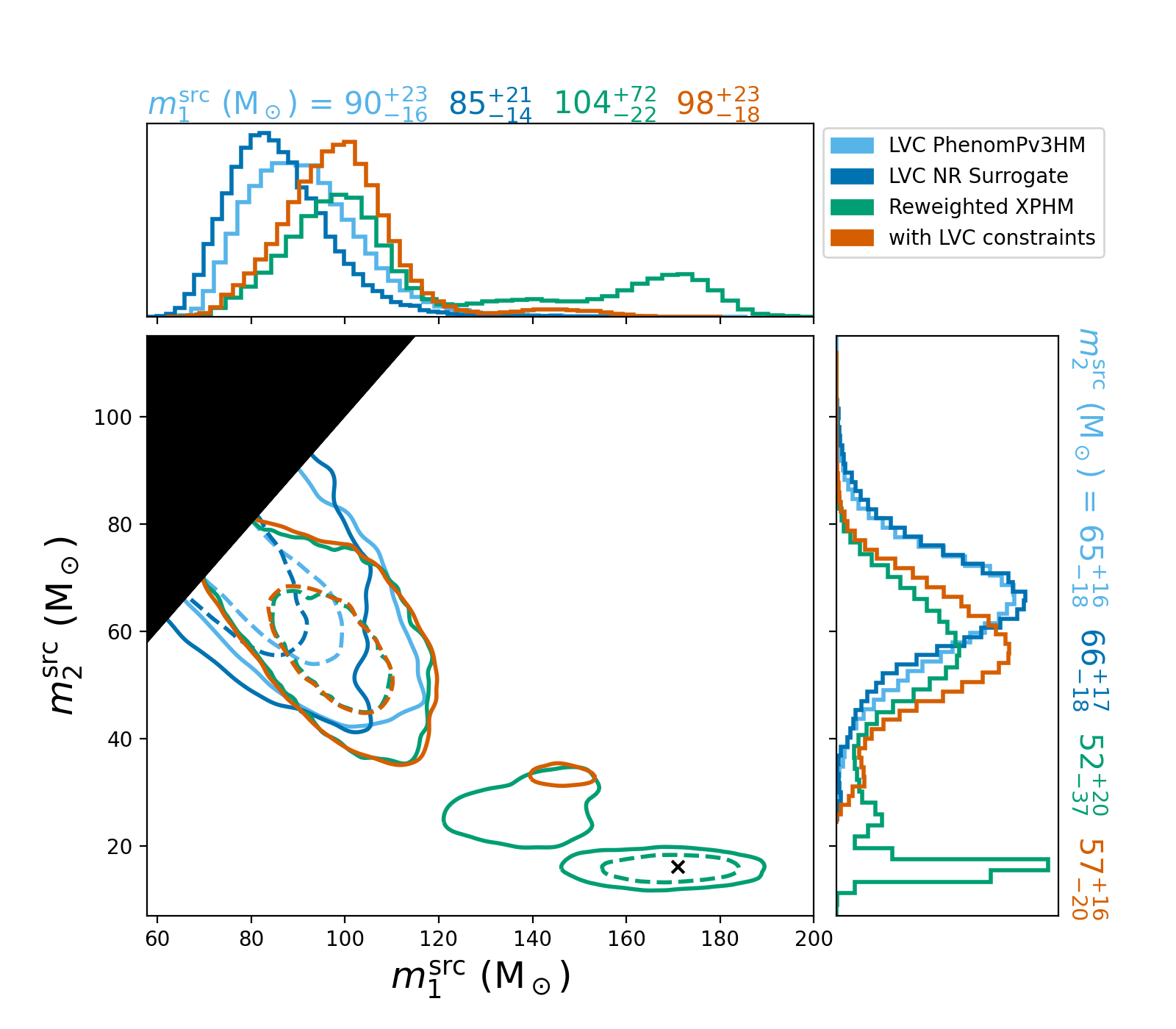}
    \caption{Comparison between our reweighted posteriors and posteriors published by the LIGO and Virgo collaborations (LVC) in \cite{Abbott:2020tfl}. The left column shows detector-frame masses; right column shows source-frame masses. The top row shows the result of reweighting our samples from a prior uniform in source-frame total mass and mass ratio to uniform in source-frame component masses. The bottom row shows the result when we additionally reweight from a prior uniform in comoving volume and source-frame masses to a prior uniform in the cube of the luminosity distance and uniform in detector-frame masses; this is the prior most similar to what was used in the LVC analysis. The median and 90\% credible interval on each parameter is reported above and to the side of the marginal distributions. Grayed regions indicate constraints used in the LVC analysis. The dotted, dashed, and solid boundaries correspond to the chirp mass, total mass, and mass ratio constraints that were used. The combination of the constraints excludes the region in which we find the maximum-likelihood waveform (black cross), and which has the largest posterior support assuming a prior uniform in mass ratio, source-frame total mass, and comoving volume.}
\label{fig:lvc}
\end{figure*}

We compare our results to those obtained in the original analysis~\citep{Abbott:2020uma} performed by the LVC using the NRSur7dq4 and IMRPhenomPv3HM waveform models in Fig~\ref{fig:lvc}; IMRPhenomXPHM was not available until well after the discovery of GW190521. If we compare our posterior using a prior that is uniform in source-frame component masses we find that there is still significant support ($51\%$) for $q>4$. We further reweight from our uniform in comoving volume prior to one which is uniform in the cube of the luminosity distance and uniform in detector-frame masses, which is most similar to that used by the LVC. We still find a second mode at high mass ratio under this prior. The LVC analysis also applied strict constraints on the prior space (shown by the shaded areas) which prevents sampling the regions of parameter space we find with highest likelihood. Combined with the resampling efficiency, we find that our results are compatible with the LVC results only after reweighting the distance prior and also applying the constraints in the LVC analysis. 

There is a clear multi-modal structure visible in our posteriors. We compare the gravitational waveform from the two dominant modes in the GW-only IMRPhenomXPHM analysis in Fig.~\ref{fig:waveform}. As expected, we see that the waveforms are similar in both modes. Each waveform is a sum of a spin-weighted spherical harmonics. To date, only two events have had measurable sub-dominant harmonics, GW190412~\citep{GW190412} and GW190814~\citep{GW190814}. If we just use the dominant harmonic in the waveform model, the signal-to-noise ratio is 14.5 and 14.7 for the low and high-mass-ratio waveforms, respectively. The total signal-to-noise ratio for the two waveforms, however, is 15.4 and 16.2, respectively. This means that nearly all of the additional signal-to-noise ratio of the maximum-likelihood waveform comes from the sub-dominant gravitational-wave harmonics.

The original LVC analysis of GW190521 used a lower-frequency cutoff of 11 Hz. We find no significant change in the likelihoods of our posterior samples if we extend our analysis down from 20 to 11 Hz. To further validate the qualitative features of the posteriors, we also performed an analysis with a simulated gravitational-wave signal whose parameters are drawn from our posterior. We find a similar multi-modal structure in the posterior for the simulated signal.

\section{Discussion}

We have analyzed GW190521 using a prior uniform in mass ratio and total mass, as well as a using a prior uniform in source-frame component measses. We find that when considering the GW data alone the component masses are consistent with a source with $q > 4$ at $93\%$ (51\%) probability for the prior uniform in $q$, $M^{\rm src}$ ($m^{\mathrm{src}}_{1,2}$). Given the likelihood of
the existence of a mass gap, which would disfavor the lower-mass-ratio mode,
we surmise that GW190521 is likely the first observed IMRI.

We also find that the reality of the flare association significantly impacts the parameter distribution. As suggested in~\cite{Graham:2020gwr} this ambiguity may be resolved if a recurrence of the flare is observed at the predicted time. Preliminary estimates of the association probability do not provide significant support nor reject the possibility based on the spatial coincidence alone~\citep{Ashton:2020kyr}. We compare the evidence reported by our flare-location constrained and agnostic analyses to estimate the likelihood of an association. We find that the odds of a common spatial origin is $\ln\mathcal{B} \gtrsim -4$ when we assume that a region of prior volume given by the low-latency localization volumes was the primary focus of follow-up observation. An optimistic upper bound can be obtained if assume that the entire prior volume we consider had been equally surveyed by ZTF and no other EM flares observed out to z$\sim1.33$. However, even under this assumption, we only find $\ln\mathcal{B}\sim 2.3$ in favor of the association.

Based on the gravitational-wave data alone, the component masses of GW190521 place both black holes outside the ``mass gap'' between $~50-120 \msun$ at the $93\%$ or $52\%$ credible level if using either a uniform in $q$, $M^{\rm src}$ or $m^{\mathrm{src}}_{1,2}$ prior, respectively. This suggests that a hierarchical merger scenario may not be required to explain GW190521. Another study has similarly suggested this based on a reweighting of the public results~\citep{Fishbach:2020qag}. However, our analysis significantly increases the support for this scenario, as we find additional modes in the posterior missed by earlier analyses. For the high-mass-ratio mode, the merger is preferentially precessing with significant spin on the primary mass $\chi_1 > 0.7$ at the 90\% credible level. Given the preference for spin anti-aligned with the orbital angular momentum, $\chi_{\mathrm{eff}} < 0$ at 97\% probability, this may suggest a dynamical capture formation scenario~\citep{Rodriguez:2016vmx,Postnov:2017nfw,Bavera:2019fkg,Safarzadeh:2020jsc}.

A limitation of all current analyses of GW190521 is the lack of a complete description of all physical effects in a single waveform model. Several analyses have suggested that GW190521 may also be consistent with an eccentric merger~\citep{Gayathri:2020coq,Romero-Shaw:2020thy,CalderonBustillo:2020odh}. At this time no model exists that includes sub-dominant harmonics, precession, eccentricity, nearly extremal spins, and support for high mass ratios simultaneously. To achieve a complete understanding of GW190521 and mitigate systematic effects, it is necessary for the numerical relativity and modelling community to work towards such a comprehensive prescription.

\section*{ACKNOWLEDGMENTS}
We make available posterior samples from our analyses along with the configuration files necessary to reproduce our results at \url{http://github.com/gwastro/gw190521}.

 We thank Thomas Dent for his comments along with Frank Ohme and Sebastian Khan for their insight into waveform modelling. We acknowledge the Max Planck Gesellschaft. We thank the computing team from AEI Hannover for their significant technical support with special thanks to Carsten Aulbert and Henning Fehrmann. This research has made use of data from the Gravitational Wave Open Science Center (https://www.gw-openscience.org), a service of LIGO Laboratory, the LIGO Scientific Collaboration and the Virgo Collaboration. LIGO is funded by the U.S. National Science Foundation. Virgo is funded by the French Centre National de Recherche Scientifique (CNRS), the Italian Istituto Nazionale della Fisica Nucleare (INFN) and the Dutch Nikhef, with contributions by Polish and Hungarian institutes. 
 
\pagebreak
\bibliography{references}

\end{document}